\documentclass[fleqn,usenatbib]{mnras}

\usepackage{newtxtext,newtxmath}

\usepackage[T1]{fontenc}
\usepackage{xcolor}

\DeclareRobustCommand{\VAN}[3]{#2}
\let\VANthebibliography\thebibliography
\def\thebibliography{\DeclareRobustCommand{\VAN}[3]{##3}\VANthebibliography}

\usepackage{graphicx} 
\usepackage{amsmath} 
\usepackage{caption}
\usepackage{subfig}
\usepackage{multirow}
\usepackage{comment}

\title[Jet proper motion in two blazars at $z>3$]{Proper motion of the radio jets in two blazars at redshift above 3}

\author[Krezinger et al.]{
Máté Krezinger,$^{1,2,3}$\thanks{E-mail: krezinger.mate@csfk.org}
Sándor Frey,$^{2,3,4}$
Krisztina Perger,$^{2,3}$
Krisztina É. Gabányi,$^{1,2,3,4,5}$
Tao An,$^{6}$
\newauthor Yingkang Zhang,$^{6}$
Leonid I. Gurvits,$^{7,8}$
Oleg Titov,$^{9}$
Alexey Melnikov,$^{10}$
Zsolt Paragi$^{7}$
\\
$^{1}$Department of Astronomy, Institute of Physics and Astronomy, ELTE E\"otv\"os Lor\'and University, P\'azm\'any P\'eter s\'et\'any 1/A, H-1117 Budapest, Hungary\\
$^{2}$Konkoly Observatory, HUN-REN Research Centre for Astronomy and Earth Sciences, Konkoly Thege Mikl\'{o}s \'{u}t 15-17, 1121 Budapest, Hungary\\
$^{3}$CSFK, MTA Centre of Excellence, Konkoly Thege Mikl\'os \'ut 15-17, H-1121 Budapest, Hungary\\
$^{4}$Institute of Physics and Astronomy, ELTE E\"otv\"os Lor\'and University, P\'azm\'any P\'eter s\'et\'any 1/A, H-1117 Budapest, Hungary\\
$^{5}$HUN-REN--ELTE Extragalactic Astrophysics Research Group, E\"otv\"os Lor\'and University, P\'azm\'any P\'eter s\'et\'any 1/A, H-1117 Budapest, Hungary\\
$^{6}$Shanghai Astronomical Observatory, Key Laboratory of Radio Astronomy and Technology, Chinese Academy of Sciences, 80 Nandan Road, Shanghai 200030, China\\
$^{7}$Joint Institute for VLBI ERIC, Oude Hoogeveensedijk 4, 7991 PD Dwingeloo, The Netherlands\\
$^{8}$Faculty of Aerospace Engineering, Delft University of Technology, Kluyverweg 1, 2629 HS Delft, The Netherlands\\
$^{9}$Geoscience Australia, PO Box 378, Canberra 2601, Australia\\
$^{10}$Institute of Applied Astronomy, Russian Academy of Sciences, Kutuzova Embankment 10, St. Petersburg, 191187, Russia
}

\date{Accepted XXX. Received YYY; in original form ZZZ}

\pubyear{2023}

\begin{document}
\label{firstpage}
\pagerange{\pageref{firstpage}--\pageref{lastpage}}
\maketitle

\begin{abstract}
There is still a limited number of high-redshift ($z>3$) active galactic nuclei (AGN) whose jet kinematics have been studied with very long baseline interferometry (VLBI). Without a dedicated proper motion survey, regularly conducted astrometric VLBI observations of bright radio-emitting AGN with sensitive arrays can be utilized to follow changes in the jets, by means of high-resolution imaging and brightness distribution modeling. Here we present a first-time VLBI jet kinematic study of NVSS~J080518$+$614423 ($z = 3.033$) and NVSS~J165844$-$073918 ($z = 3.742$), two flat-spectrum radio quasars that display milliarcsecond-scale jet morphology. Archival astrometric observations carried out mainly with the Very Long Baseline Array, supplemented by recent data taken with the European VLBI Network, allowed us to monitor changes in their radio structure in the $7.6-8.6$~GHz frequency band, covering almost two decades. By identifying individual jet components at each epoch, we were able to determine the apparent proper motion for multiple features in both sources. Apparent superluminal motions range between $(1-14)\,c$, and are found to be consistent with studies of other high-redshift AGN targets. Using the physical parameters derived from the brightness distribution modeling, we estimate the Doppler-boosting factors ($\delta \approx 11.2$ and $\delta \approx 2.7$), the Lorentz factors ($\Gamma \approx 7.4$ and $\Gamma \approx 36.6$) and the jet viewing angles ($\theta \approx 4\fdg4$ and $\theta \approx 8\fdg0$), for NVSS~J080518$+$614423 and NVSS~J165844$-$073918, respectively. The data revealed a stationary jet component with negligible apparent proper motion in NVSS~J165844$-$073918.
\end{abstract}




\section{Introduction} 
\label{sec:introduction} 

Active galactic nuclei (AGN) can be found and studied through almost the entire history of the Universe. These are the most powerful non-transient objects we currently know and can be observed even at extremely high redshifts \citep[$z>7$, e.g.][]{2018Natur.553..473B, 2021ApJ...907L...1W,2023ApJ...948L..14C,2023ApJ...953L..29L}. AGN provide essential information about the behaviour of supermassive black holes (SMBHs) and the connection to their host galaxies. They are powered by accretion onto their central SMBHs, thus releasing a large amount of energy across the whole electromagnetic spectrum. The most massive high-redshift SMBHs known reach masses of $\sim 10^{10}\,{\rm M_{\odot}}$ \citep[e.g.][]{2013ApJ...773...44W,2019ApJ...873...35S,2021Galax...9...23S}. The rapid growth of SMBHs in the early history of the Universe is one of the most intriguing fields of research in modern astrophysics and cosmology (see e.g. \citealt{2022arXiv221206907F} and references therein). The early accretion of AGN can be explained by maintaining powerful relativistic jets that carry off a significant fraction of the released gravitational energy (\citealt{2008MNRAS.386..989J, 2013MNRAS.432.2818G}). This way a black hole can keep a high accretion rate for a long time, which means faster mass growth. It is also possible for the jets to trigger an infall of galactic matter in the vicinity of the black hole \citep{2012ARAA..50..455F}. Such feedback has already been observed in local active galaxies (e.g \citealt{1998AA...331L...1S, 2018NatAs...2..179C, 2018ApJ...857..121Y}). High-redshift jetted AGN more likely harbour black holes with mass exceeding $10^9$\,M$_{\rm \odot}$ \citep{2022A&A...663A.147S}, suggesting that jets indeed play an important role in the fast formation of the first SMBHs. They also contribute to the AGN--host galaxy feedback \citep{2012ARAA..50..455F}.

The number of the AGN found at high redshifts has been increasing steadily over the years (e.g. \citealt{2016ApJ...833..222J, 2019ApJ...884...30W,2022MNRAS.511..572O,2023arXiv230606308D,2023AJ....165..191Y}). Only a small fraction ($\sim10\%$) of these quasars are radio-loud, i.e., have powerful relativistic jets producing synchrotron radio emission. \citet{2017FrASS...4....9P} showed that above $z\gtrsim 4$, there are only about $300$ AGN with radio emission above a typical sensitivity threshold of modern radio telescopes. The majority of them are weak radio sources with mJy-level flux densities at GHz frequencies. At high redshifts, radio-loud AGN are mainly represented by blazars \citep[e.g.][]{2022A&A...663A.147S}.

Blazars are a subclass of AGN in which the jet viewing angle is close ($\theta \lesssim 10\degr$) to the line of sight \citep{1995PASP..107..803U}. Due to relativistic beaming effects, they tend to have Doppler-boosted radio emission and extremely high brightness temperatures. These objects are often luminous X-ray and/or $\gamma$-ray sources, but less frequently found with increasing redshift (e.g. \citealt{2013MNRAS.433.2182S, 2015MNRAS.448.1060G}). At high redshifts and with high angular resolution, blazars are usually characterised by a compact, flat-spectrum core or a core-dominated, moderately extended jet structure (e.g. \citealt{2016MNRAS.463.3260C,2017MNRAS.467..950C,  2022ApJS..260...49K}). Apart from the jet base (the core), jet components have steep spectrum, i.e. their brightness rapidly decreases as the frequency increases. The rest-frame frequency ($\nu_0$), where the radiation is emitted, is a function of the observed frequency ($\nu_{\rm obs}$) and the redshift as $\nu_0 = (1+z)\,\nu_{\rm obs}$. Thus, at a given $\nu_{\rm obs}$, the higher the redshift, the fainter is the steep-spectrum jet (e.g. \citealt{2000pras.conf..183G, 2015IAUS..313..327G}). This effect causes the steep-spectrum jets with extended radio emission to remain largely undetected, therefore it is the cores of flat-spectrum radio quasars \citep[FSRQs,][]{2007ApJS..171...61H} with strong, compact, enhanced synchrotron emission which are typically observed at high redshifts. Milliarcsecond (mas) scale jets of high-redshift blazars tend to appear shorter than jets in the low-redshift Universe. Apart from the observational effect explained above, this is possibly also because of an intrinsic phenomenon: the dense circumnuclear matter that hinders the development of large-scale jets, as only the most powerful ones are able to penetrate it (\citealt{2016MNRAS.461L..21G, 2022MNRAS.511.4572A}). Large-scale emission structures are also dimmer due to the interaction of their electrons with photons of the cosmic microwave background \citep{2015MNRAS.452.3457G}. To detect the radio emission coming from the innermost regions of these distant sources, high angular resolution and high sensitivity are needed. The technique of very long baseline interferometry (VLBI) is capable of achieving mas-scale angular resolution which corresponds to pc scales in linear resolution.

To study the kinematics of the jets, and to acquire key information for understanding the nature of their function, VLBI is essential. With long-term VLBI monitoring, the proper motion of the jet components can be directly measured. From the derived apparent brightness temperature values, we can calculate the properties of the relativistically enhanced plasma, i.e. the Doppler-boosting factor, the bulk Lorentz factor, and the jet inclination angle. This information can be used to further refine the modeling of the spectral energy distribution (SED, e.g. \citealt{2022A&A...663A.147S}). Existing high-redshift jet proper motion studies are still scarce and usually only deal with a few sources at a time (e.g. \citealt{2010A&A...521A...6V, 2015MNRAS.446.2921F, 2018MNRAS.477.1065P, 2020MNRAS.497.2260A, 2020SciBu..65..525Z, 2022ApJ...937...19Z}). Jet component proper motion is measurable for core--jet type sources, and these are relatively rare at high redshifts. Without a dedicated long-term VLBI observing programme for determining the proper motion of high-redshift jets, a straightforward way is to look for frequently observed sources with core--jet characteristics suitable for component identifications across the epochs.

While only about 20 sources with jet proper motion analysis exist at $z > 3$, they provide important constraints on early SMBH growth and relativistic jet properties in the early Universe. Even incremental increases in sample size can improve the statistical constraints on the physics of high-redshift jets. To expand this limited sample in order to work towards improving the statistical constraints, we specifically targeted two FSRQs, NVSS~J080518$+$614423 (J0805$+$6144 in short, $z = 3.033$, \citealt{2005ApJ...626...95S}) and NVSS~J165844$-$073918 (J1658$-$0739, $z = 3.742$, \citealt{2008ApJS..175...97H}) because they are among the brightest high-redshift blazars accessible to VLBI imaging. Their flat radio spectra and compact mas-scale core--jet structures made them suitable for multi-epoch model-fitting analysis to measure jet kinematics. Both can be found among the targets of geodetic and astrometric VLBI observations \citep{2017JGeod..91..711N}, but they have not been the subjects of dedicated VLBI monitoring.

J0805$+$6144, with coordinates in the 3rd realization of the International Celestial Reference Frame (ICRF3, \citealt{2020A&A...644A.159C}) right ascension $\alpha_\mathrm{VLBI}=08^\mathrm{h}05^\mathrm{m}18\fs179562$ and declination $\delta_\mathrm{VLBI}=61\degr44\arcmin23\farcs70056$, is also known as a $\gamma$-ray source in the \textit{Fermi} Large Area Telescope (LAT) catalogues \citep{2010ApJS..188..405A,2015ApJS..218...23A}. Significant $\gamma$-ray variability is found by \citet{2012ApJS..199...31N}, and \citet{2018ApJ...853..159L} identified two flaring events, one in 2009 January, showing intra-day $\gamma$-ray variability, and another one in 2010 January. Contrary to the $\gamma$-ray variability of the object, the \textit{Swift} X-ray Telescope (XRT) measured constant X-ray flux during the first ten years of \textit{Fermi}-LAT operation, from 2008 August to 2018 August \citep{2020MNRAS.498.2594S}. Several Doppler-factor estimates can be found in the literature, based on multiwavelength SED fitting. \citet{2018ApJS..235...39C} derived $\delta = 6.4$, also giving an estimate on the central black hole mass $10^{9.1}$\,M$_{\odot}$. Later, \cite{2020MNRAS.498.2594S} found $\delta = 14.0 \pm 0.6$, while \citet{2020ApJS..248...27T} obtained a value higher by a factor of $\sim2$, $\delta = 28.8$. \citet{2020ApJ...897..177P} estimated $\delta = 22.6$, while also giving a Lorentz-factor $\Gamma = 14$. In contrast, \citet{2020MNRAS.498.2594S} found $\Gamma =1.13 \pm 0.10$, however, based only on a single X-ray observation. J0805$+$6144 is bright enough in optical to be detected by the \textit{Gaia} astrometric space telescope (\citealt{2016A&A...595A...1G, 2021A&A...649A...1G}). 

J1658$-$0739 (ICRF3 position: $\alpha_\mathrm{VLBI}=16^\mathrm{h}58^\mathrm{m}44\fs061966$, $\delta_\mathrm{VLBI}=-07\degr39\arcmin17\farcs6945$) is also a regularly observed radio source. Its NVSS 1.4-GHz flux density is ($0.83\pm0.03$)~Jy. Its FSRQ classification was confirmed by \citet{2007ApJS..171...61H}. The source is considered as a $\gamma$-ray-quiet blazar \citep{2017ApJ...851...33P}. At high energies, it has only a single X-ray detection by the \textit{Swift}-XRT instrument \citep{2017ApJ...851...33P}. With SED fitting, the Doppler factor and the bulk Lorentz factor are estimated as $\delta = 15.7$ and $\Gamma=10$, respectively. They found that the jet has an inclination angle of $3\fdg0$ and the black hole mass is estimated to be $10^{9.48}$\,M$_{\odot}$. In optical, the source is detected by \textit{Gaia}. J1658$-$0739 was imaged in the \textit{VSOP} (VLBI Space Observatory Programme) Prelaunch Survey with the VLBA at 5~GHz \citep{2000ApJS..131...95F}. The image shows a bright core with jet extending to the northeastern direction. However, later the source was not observed on ground--space interferometer baselines in the \textit{VSOP} AGN Survey \citep{2000PASJ...52..997H}.

In this work, we present jet proper motion studies of the two high-redshift blazars, J0805$+$6144 and J1658$-$0739, using VLBI measurements spanning almost two decades, based on data from new and archival observations. Such analysis is carried out for the first time for these two radio quasars, increasing the size of the $z>3$ proper motion sample by $\sim 10$ per cent. Combining the multi-epoch VLBI observations with the available radio spectral information and the \textit{Gaia} Data Release 3 (DR3) positions, we investigate the nature of the sources by determining their jet proper motions and constraining key physical parameters such as Doppler and Lorentz factors. We aim to add new data points to the sparsely sampled $z > 3$ region of the apparent proper motion--redshift relation. The results will help refine our understanding of the typical velocities and orientations of the most relativistic jets during the era of early black hole growth.
Throughout this paper, we assume a standard $\Lambda$CDM cosmological model with $\Omega_{\rm m} = 0.3$, $\Omega_{\Lambda} = 0.7$, and $H_{0} = 70$~km\,s$^{-1}$\,Mpc$^{-1}$. To determine the projected linear sizes and luminosity distances, we used the cosmology calculator of \citet{Wright_2006}.

\section{New VLBI observations and archival data}
\label{sec:obs}

J0805$+$6144 and J1658$-$0739 were observed as part of a sample of $z > 3$ blazars with the European VLBI Network (EVN) under the project code ET036 (PI: O. Titov). Altogether, three epochs of observations were made, on 2018 June 4 (segment ET036A), 2018 October 27 (ET036B), and 2019 February 25 (ET036C). The primary purpose of the project was astrometric in nature, to investigate the positional stability of prominent high-redshift radio sources \citep{2023AJ....165...69T}. However, source brightness distribution maps were also made from the interferometer visibility data. The astrometric-style dual-band snapshot observations were performed at 2.3~GHz (S band) and 8.6~GHz (X band). The data were recorded in right circular polarization and included a total of 16 intermediate frequency channels (IFs), 6 in S band and 10 in X band, each 8-MHz wide. The total bandwidth was 128~MHz and the data rate was $512$\,Mbps. Table~\ref{tb:observations} contains the details of the EVN observations, including the dates, frequencies, target sources, and the participating telescopes in each project segment. The observed data were processed at the Joint Institute for VLBI European Research Infrastructure Consortium (JIVE, Dwingeloo, The Netherlands) with the SFXC software correlator \citep{2015ExA....39..259K} with $0.5$~s integration time and $62.5$~kHz spectral resolution. 

\begin{table*}
     \centering
    \caption{Details of observations used from the EVN project ET036.}
    \begin{tabular}{ccccccc}
    \hline
Project & Frequency & Observing date & Participating radio telescopes & Source & Number of & Total on-source \\ 
segment & $\nu_{\rm obs}$ [GHz] & [yyyy-mm-dd] & & & scans & time [min] \\
(1)    & (2)       & (3)             & (4)             & (5)     & (6)        & (7) \\
    \hline
 ET036A  & 8.6 & 2018-07-04 & Ef, Mc, Nt, O6, Ys, Sv, Zc, Bd & J0805$+$6144 & 4 & 44.6 \\
         &     &            & Ef, Mc, Nt, O6, Ys, Sv, Zc, Hh & J1658$-$0739 & 2 & 2.9  \\
 ET036B  & 8.6 & 2018-10-27 & Ef, Mc, O6, T6, Ys, Sv, Zc, Bd, Wz & J0805$+$6144 & 4 & 4.6 \\
         &     &             & Ef, Mc, O6, Ys, Hh, Zc, Wz & J1658$-$0739 & 3 & 2.2 \\
 ET036C  & 8.6 & 2019-02-25 & Ef, Mc, O6, T6, Ur, Ys, Sv, Zc, Bd, Wn & J0805$+$6144 & 5 & 6.4 \\
         &     &             & Ef, Mc, O6, T6, Ur, Ys, Hh, Sv, Zc, Bd, Wn & J1658$-$0739 & 3 & 46.3 \\
    \hline
    \end{tabular}
    \\
    \noindent Notes: Radio telescope codes: Effelsberg (Ef, Germany), Medicina (Mc, Italy), Noto (Nt, Italy), Onsala 20-m (O6, Sweden), Tianma (T6, China), Urumqi (Ur, China), Yebes (Ys, Spain), Svetloe (Sv, Russia), Zelenchukskaya (Zc, Russia), Badary (Bd, Russia), Hartebeesthoek (Hh, South Africa), Wettzell 20-m (Wz, Germany), Wettzell 13-m North (Wn, Germany).
    \label{tb:observations}
\end{table*}
\begin{table*}
    \centering
    \caption{Details of observations of the archival data obtained from the Astrogeo database 
    }
    \begin{tabular}{ccccccccc}
    \hline
Source & Project & Frequency & \multicolumn{2}{c}{Observing date} & Participating VLBI  &  On-source & Bandwidth & Reference
\\ & code & $\nu_{\rm obs}$ [GHz] & [yyyy-mm-dd]  & [MJD] & antennas & time [min] & [MHz]  &  \\
(1)    & (2)       & (3)             & (4)             & (5)     & (6)        & (7) & (8) \\
    \hline
J0800$+$6144 & bf071 & 8.6 & 2002-05-14 & 52408 & VLBA & 5.2 & 8 $\times$ 4 & \citet{2003AJ....126.2562F} \\
             & bk124 & 8.6 & 2005-07-09 & 53560 & VLBA & 1.5 & 8 $\times$ 4 & \citet{2007AJ....133.1236K} \\
             & rdv76 & 8.6 & 2009-07-29  & 55041 & VLBA + Kk, Ny, Wz, Zc & 130 & 8 $\times$ 4 &  \\
             & bc191c(3) & 8.6 & 2010-08-03 & 55411 & VLBA  & 8.4 & 16 $\times$ 8 &  \\ 
             & s3111a & 8.6 & 2010-12-05 & 55535 & VLBA & 1 & 16 $\times$ 8 &  \\
    \hline
J1658$-$0739 & bf071 & 8.4 & 2002-01-31 & 52305 & VLBA & 6.2 & 8 $\times$ 4 & \citet{2003AJ....126.2562F} \\
             & rdv72 & 8.6 & 2008-12-17 & 54817 & VLBA + Kk, Wz, Zc & 76 & 16 $\times$ 4 &  \\ 
             & rdv76 & 8.6 & 2009-07-29 & 55041 & VLBA + Kk, Wz, Zc & 12 & 8 $\times$ 4 &  \\
             & bc196p(3) & 8.4 & 2011-05-15 & 55696 & VLBA & 27 & 16 $\times$ 8 &  \\
             & bc196zn(3) & 8.4 & 2011-12-19 & 55914 & VLBA & 18.8 & 16 $\times$ 8 &  \\
    \hline
    \end{tabular}
    \\
    \noindent Notes: We only present here several rows as examples, the table is available in its entirety in machine-readable form. The columns are as follows: Col.~1 -- Source name; Col.~2 -- Project code; Col.~3 -- Observing frequency; Col.~4 -- Observing date; Col.~5 -- Observing date in MJD; Col.~6 -- Codes of participating radio telescopes. VLBA: Very Long Baseline Array. Additional antennas: Fortaleza (Ft, Brazil), Hartebeesthoek (Hh, South Africa), Kokee Park (Kk, Hawaii, USA), Onsala (On, Sweden), Matera (Ma, Italy), Ny {\AA}lesund (Ny, Norway), Westford (Wf, USA), Wettzell 13-m North (Wn, Germany), Wettzell (Wz, Germany), Yarragadee 12-m (Yg, Australia), Zelenchukskaya (Zc, Russia); Col.~7 -- Sum of the scan durations in minutes; Col.~8 -- Observing bandwidth of each IF (in MHz) times the number of IFs; Col.~9 -- Literature reference where available.
    \label{tb:astrogeo}
\end{table*}

Both J0805$+$6144 and J1658$-$0739 are frequently observed blazars and have VLBI data available in the archives covering almost two decades. We collected the archival calibrated 7.6$-$8.6 GHz (X band) VLBI visibility data from the Astrogeo\footnote{\url{http://astrogeo.org/vlbi_images/}} database. These observations were usually performed in the framework of VLBI calibrator surveys and astrometric/geodetic VLBI experiments (\citealt{2003AJ....126.2562F, 2007AJ....133.1236K, 2016AJ....151..154G, 2017ApJ...838..139S, 2021AJ....162..121H}). The sources were mostly observed with the VLBA, occasionally supplemented with other radio telescopes, in snapshot mode at S/X bands. Such observations are usually scheduled with few-minute scans on the targets, rapidly alternating between objects seen in different directions \citep[e.g.][]{1998RvMP...70.1393S}. For imaging purposes, snapshots widely spaced in time could provide good $(u,v)$ sampling. The time coverage of archival VLBI data is not uniform, the observations are more frequent after 2017. Table~\ref{tb:astrogeo} summarises the archival calibrated visibility data we downloaded and imaged. Even taking the cosmological time dilation into account, the $\sim 20$ years in the observer's frame still translate to 4 years in the rest-frame of J0805$+$6144 and J1658$-$0739, giving sufficiently long time to detect the apparent proper motion of jet components. 

\section{Data reduction}
\label{sec:data}

\subsection{EVN observations}
\label{subsec:data_evn}

We calibrated the EVN data using the U.S National Radio Astronomy Observatory (NRAO) Astronomical Image Processing System (\texttt{AIPS}) software package \citep{2003ASSL..285..109G}, following a standard procedure (e.g. \citealt{1995ASPC...82..227D}). The interferometric visibility amplitudes were calibrated using the antenna gain curves and the system temperatures were measured at the telescopes. For some antennas, only nominal system temperature values were available (ET036A: Zc, T6; ET036B: Zc, Sv, Ur, Wz; ET036C: Zc, Sv, Ur, Wn; for station codes, see Table~\ref{tb:observations}). Ionospheric delays were obtained from global navigation satellite systems measurements, and the phases were corrected using the measured Earth orientation parameters. As a simple bandpass correction, we flagged the first and last 4 channels in each IF. Short 1-min data scans were chosen to solve for instrumental phases and delays. Global fringe-fitting \citep{1983AJ.....88..688S} was performed on the target sources and the bright fringe-finder sources were also scheduled in the experiments. Finally, the calibrated data were averaged in frequency. 

The integration times spent on the sources in each observing session were markedly different (Table~\ref{tb:observations}), which resulted in poor image quality in 2 out of the 3 observing epochs. Since the epochs were close in time, only spanning a few months in the rest frame of the sources, it was feasible to combine the calibrated visibility data in \texttt{AIPS}. By doing so, it was possible to reduce the noise level in the final images, while keeping all data. We associated the date to the combined observations when the on-source time was much longer than at the other two epochs. It is 2018 Jul 4 for J0805$+$6144 and 2019 Feb 25 for J1658$-$0739. 

The combined calibrated data of the target sources were exported from \texttt{AIPS} for further work in \texttt{Difmap} \citep{1994BAAS...26..987S}. There we performed standard hybrid mapping to produce the images of the sources, which includes iterations of the \texttt{CLEAN} algorithm \citep{1974A&AS...15..417H} and phase-only self-calibration \citep{1984ARA&A..22...97P} followed by a few rounds of amplitude and phase self-calibration. Gaussian brightness distribution model components were fitted to the self-calibrated visibility data \citep{1995ASPC...82..267P}, to quantitatively characterize sizes and flux densities of the core and jet components. The core of each source was fitted with elliptical Gaussian model components, while the jet components with circular Gaussians. Elliptical Gaussians fitted to the core provide directional information on the innermost, barely resolved section of the jet. This way we trace possible changes in the position angle of the major axis. 

\subsection{Archival data}
\label{subsec:data_archival}

After downloading all the available calibrated and self-calibrated X-band visibility data for J0805+6144 and J1658$-$0739 from the Astrogeo archive, we made images and modelfits by following the procedure as described above. Repeating self-calibration on already self-calibrated data cannot cause any bias. In turn, it may slightly improve the quality with respect to the automatically-generated Astrogeo images, especially in the case of a complex source structure and if outlier visiblity points are flagged manually. We note that, similarly to what is done with our EVN data, many of the archival Astrogeo visibility data sets are combined, containing observations made by different subarrays and at epochs typically separated by hours to days. Two imaging methods were used to ensure consistency in the resulting images and model parameters. First, the data were processed using an automatic \texttt{Difmap} imaging script\footnote{\url{https://github.com/rstofi/VLBI_Imaging_Script} \citep{2017IAUS..324..247R}}, to quickly obtain an initial image of the two sources at each epoch. Then we also imaged data from all the epochs manually in \texttt{Difmap}. Self-calibrated visibility data obtained after hybrid mapping served as a basis for model fitting using Gaussian brightness distribution components. Similarly, the core components were fitted with elliptical Gaussians, except for a few cases where a circular Gaussian component provided a better fit. The jet components further away from the core were fitted with circular Gaussians. At most of the epochs, two jet components were found, mainly the two closest ones to the core. At a few epochs, when the array configuration and higher sensitivity allowed imaging of some extended emission, outer jet components also appeared. For epochs close to each other in time, common starting models were used to ensure the consistency of the results.

\subsection{Modelfit uncertainties}  

The errors of the fitted model component parameters (except for the separation parameter) were estimated following \citet{1999ASPC..180..301F}. The method considers the statistical error in the image. We applied an additional 10 per cent absolute amplitude calibration uncertainty for the fitted flux density of each component (e.g. \citealt{2011A&A...526A..74M,2012A&A...544A..34P,2018MNRAS.473.1388M,2021ApJ...919...40P}). The uncertainties of the component separation were found to be underestimated (with up to $\sim3$ orders of magnitude smaller errors compared to the separation values themselves) if estimated following \citet{1999ASPC..180..301F}. To tackle this problem, we considered 20 per cent of the full width at half-maximum (FWHM) of the synthesised beam along the component position angle (e.g. \citealt{2009AJ....138.1874L,2013AJ....146..120L,2021ApJ...919...40P}) as the separation uncertainty. For snapshot observations typically with short observing time and sparse $(u,v)$ coverage, this conservative estimate gives more reasonable results. Table \ref{tb:results} contains the modelfit results with their calculated uncertainties.

\begin{table*}
    \centering
    \caption{Model-fitting parameters and the calculated physical properties}
    \label{tb:results}
    \begin{tabular}{ccccccccc}
    \hline
Source & Epoch & Component & $S_{\nu}$ & $R$ & $\phi$ & PA & $T_{\rm b}$ & PA$_{\rm core}$
\\  &  &  & [Jy] & [mas] & [mas] & [$\degr$] & [$\times 10^{10}$ K] & [$\degr$] \\
(1)    & (2)       & (3)             & (4)             & (5)     & (6)        & (7) & (8) & (9) \\
    \hline
J0800$+$6144 & 2002-05-14 & C & 0.929 (0.105) & & 0.4 $\times$ 0.3 &  & 51.8 (6.0) & $-2.2$ \\
             & ...        & J1 & 0.079 (0.016) & 0.61 (0.20) & 0.5 & 191.6 (0.7) & & \\
             & ...        & J2 & 0.052 (0.014) & 1.97 (0.36) & 0.9 & 154.3 (0.3) & & \\
             & 2005-07-09 & C & 0.799 (0.096) & & 0.5 & & 26.3 (3.3) &  \\
             & ...        & J1 & 0.127 (0.024) & 0.99 (0.26) & 0.8 & 188.8 (0.5) & & \\
             & 2009-07-29 & C & 0.713 (0.085) & & 1.0 $\times$ 0.3 & & 15.4 (2.1) & $-9.8$ \\
             & ...        & J1 & 0.070 (0.013) & 1.28 (0.11) & 0.6 & 183.1 (0.1) & & \\
             & ...        & J2 & 0.014 (0.007) & 2.22 (0.12) & 0.4 & 151.4 (0.1) & & \\
             & 2010-08-03 & C & 0.419 (0.051) & & 0.8 $\times$ 0.2 & & 20.2 (2.8) & $-4.4$ \\
             & ...        & J1 & 0.079 (0.014) & 1.52 (0.29) & 1.0 & 168.4 (0.2) & & \\
    \hline
J1658$-$0739 & 2002-01-31 & C & 0.935 (0.107) & & 0.8 $\times$ 0.3 & & 35.2 (4.2) & 15.7 \\
             & ...        & J3 & 0.022 (0.010) & 3.44 (0.30) & 0.1 & 27.8 (0.1) & & \\
             & ...        & J4 & 0.010 (0.009) & 8.72 (0.33) & 0.1 & 21.9 (0.1) & & \\
             & 2008-12-17 & C & 0.595 (0.074) & & 1.7 $\times$ 0.5 & & 5.8 (0.8) & 26.5 \\
             & ...        & J1 & 0.040 (0.013) & 1.24 (0.26) & 0.1 & 47.9 (0.4) & & \\
             & 2009-07-29 & C & 0.085 (0.012) & & 0.54 $\times$ 0.01 & & > 26.0 & 51.8 \\
             & ...        & J1 & 0.460 (0.055) & 1.29 (0.17) & 0.4 & 48.4 (0.4) & & \\
             & ...        & J2 & 0.158 (0.023) & 2.28 (0.20) & 0.6 & 38.4 (0.2) & & \\
             & ...        & J3 & 0.017 (0.007) & 5.57 (0.25) & 0.8 & 25.5 (0.1) & & \\
             & ...        & J4 & 0.008 (0.007) & 10.07 (0.33) & 0.7 & 15.2 (0.1) & & \\
    \hline
    \end{tabular}
    \\
    \noindent Notes: We only present here several rows as examples, the table is available in its entirety in machine-readable form. The $1\sigma$ error is given in parentheses for the derived quantities. The columns are as follows: Col.~1 -- Source name; Col.~2 -- Observing epoch (year-month-day); Col.~3 -- Identifier of the core (C) and jet (J) components, the latter are numbered based on the increasing distance from the core; Col.~4 -- Fitted flux density of the component; Col.~5 -- Separation of the jet component from core; Col.~6 --  FWHM size of the fitted Gaussian model component; the sizes for the core components fitted with an elliptical Gaussian are written as $a \times b$, where $a$ and $b$ are the major and minor axes; Col.~7 -- Position angle of the component with respect to the core, measured from north to east;  Col.~8 -- Calculated redhsift-corrected brightness temperature; Col.~9 -- Position angle of the major axis of the fitted elliptical Gaussian core component, measured from north to east.
\end{table*}

\section{Results}
\label{sec:results}

\subsection{Radio morphology of the sources}
\label{subsec:results_morp}

Figures~\ref{fig:J0805+6144-pm} and \ref{fig:J1658-0739-pm} show examples of 8.6-GHz VLBI images taken on 2018 May 19 for J0805$+$6144 (experiment ug002h) and on 2017 Aug 1 for J1658$-$0739 (experiment rv125, Table~\ref{tb:astrogeo}). For both sources, the core--jet structures extending up to about $10$~mas are typical for blazars, although this is not unique for blazars, and has been observed in various quasars, including those at high redshifts (e.g. \citealt{2010A&A...524A..83F,2016MNRAS.463.3260C,2022ApJS..260...49K}). The \textit{Gaia} DR3 optical coordinates are marked in the radio images with red crosses whose size represents the $1\sigma$ positional errors. {Broad optical emission lines used for the measurement of the quasar redshift are known to be formed in the so-called broad-line region (BLR) in the vicinity of the central black hole \citep{2006LNP...693...77P}. If the optical emission is dominated by the non-thermal emission from the relativistic jet, the thermal optical emission from the BLR would be faded out and the broad emission lines are not detected. This is a common situation with many blazars whose redshift is unknown so far in spite of their high brightness in the optical wavelengths. Therefore the detection of the broad emission lines in the optical spectra of both quasars favours that the thermal origin of the optical emission and, correspondingly, the \textit{Gaia} position better represents the location of the central black hole than the radio core \citep{2017A&A...598L...1K,Plavin_2019}. The latter is in fact the synchrotron self-absorbed base of the jet with $\tau_{\nu}=1$ optical depth at the given frequency $\nu$. Because absolute astrometric information is lost after fringe-fitting, we associated the position of the VLBI brightness peaks with the X-band ICRF3 position of the respective quasars. The VLBI astrometric position generally reflects the location of the radio brightness peak \citep[e.g.][]{1997AJ....114.2284F}. However, according to \citet{2009A&A...505L...1P}, there could be a $\sim 0.2$-mas level offset between the phase-referenced and group-delay positions along the jet direction, caused by opacity effects on the emission at the jet base. Moreovoer, its actual value may be affected by the variable core-shift effect. Up to a few mas offsets between the radio and optical AGN positions are found to statistically coincide with the VLBI jet direction \citep{2019MNRAS.482.3023P}. Indeed, in both sources, the \textit{Gaia} optical positions are offset from the radio brightness peak that indicates the location of the core within $\sim 1$~mas, apparently upstream along the jet (Figs.~\ref{fig:J0805+6144-pm}--\ref{fig:J1658-0739-pm}). 

The 8.6-GHz image of J0805$+$6144 (Fig.~\ref{fig:J0805+6144-pm}) shows the core and two jet components (J1, J2) close to it, within $\sim 3$~mas. The innermost jet starts pointing to the south, then seems to turn eastwards at $\sim 1$~mas from the core, and continues in the southeastern direction. Further downstream, there is a third, diffuse jet component (J3) at $\sim 8$~mas separation. This feature is heavily resolved, or even remains undetected at most epochs where the imaging sensitivity is insufficient. 

The image of J1658$-$0739 (Fig.~\ref{fig:J1658-0739-pm}) shows a core and four jet components. The two inner components (J1, J2) are within $\sim2$~mas from the core. The jet points to the northeast, then gradually bends northwards. 

\subsection{Jet proper motion}
\label{subsec:proper_motion}

To estimate the apparent proper motion of jet components, the first task was to identify the same features across the epochs spanning about two decades of observations. As observations were made with a wide range of imaging sensitivity and varying angular resolution, not all epochs provided the same number of detected jet components. The apparent proper motions of securely identified jet components were calculated based on the relative positions of the fitted circular Gaussian brightness distribution models with respect to the core. The measured separations and their corresponding errors do not allow us to estimate reliably any higher time derivative of the separation than the first one. Therefore the angular separation vs. time was fitted with linear function, and the apparent proper motion is taken as the slope of this line. Figures~\ref{fig:J0805+6144-pm} and \ref{fig:J1658-0739-pm} show the core--jet component separations and position angles as a function of time, along with the best-fit linear trends for both J0805$+$6144 and J1658$-$0739. For the VLBI images shown as examples, we choose observing epochs where all fitted jet features can be seen. In Table~\ref{tab:pm}, we collect the fitted kinematic parameters, as well as geometric and physical parameters of the jets calculated in Sect.~\ref{subsec:lorentz}. Using the apparent proper motions, we can determine whether the jet is superluminal. AGN with relativistic jets pointing close to the line of sight often show superluminal motion  which is an apparent relativistic effect \citep{1966Natur.211..468R}. 

\begin{table*}
    \centering
    \caption{Kinematic and physical properties of the jets}
    \label{tab:pm}
    \begin{tabular}{c|c|c|c|c|c|c|c|c}
    \hline
    Source & Component & $\mu_{\rm r}$   & $\mu_{\rm PA}$  & $\beta$ & $T_\mathrm{b,median}$ &  $\delta$ & $\Gamma$  & $\theta$ \\
           &           & [mas\,yr$^{-1}$] & [$\degr$\, yr$^{-1}$] & [$c$]  & [10$^{11}$~K]   & &    & [$\degr$]        \\
    (1)    & (2)       & (3)             & (4)             & (5)     & (6)        & (7) & (8) & (9)\\
    \hline
    J0805$+$6144 & C & & & & 4.59 & 11.2 &  7.4 &  4.4 \\
                 & J1 & 0.044 (0.010) & $-1.461$ (0.228) & 4.5 (0.9) & &   & \\ 
                 & J2 & 0.062 (0.010) & $-0.025$ (0.089) & 6.3 (1.0) & &   & \\
                 & J3 & $<$ 0.0972 & $-0.579$ (0.545) & $<$ 9.8  & &       & \\
    \hline
    J1658$-$0739 & C & & & &  1.11 & 2.7  & 36.6 &  8.0  \\
                 & J1 & 0.032 (0.008) & $-0.019$ (0.244) & 3.5 (0.9) & &   &   \\
                 & J2 & 0.006 (0.014) & $-0.464$ (0.178) & 0.9 (1.6) & &   &   \\
                 & J3 & 0.124 (0.061) & $-0.419$ (0.136) & 13.7 (5.2) & &  &  \\
                 & J4 & 0.125 (0.066) & $-0.607$ (0.234) & 13.8 (6.4) & &  &   \\
    \hline
    \end{tabular}
    \\
    \noindent Notes: The columns are as follows: Col.~1 -- Source name; Col.~2 -- Identifier of the core (C) and jet (J) components, the latter are numbered based on their increasing distance from the core; Col.~3 -- Jet radial proper motion; Col.~4 -- Rate of change of the jet position angle; Col.~5 -- Apparent transverse radial speed in the units of the speed of light; Col.~6 -- The median brightness temperature of the core; Col.~7 -- Doppler factor calculated from the VLBI core median brightness temperature; Col.~8 -- Bulk Lorentz factor;  Col.~9 -- Jet viewing angle.
\end{table*}

For J0805$+$6144, the apparent speeds of the innermost two jet components, J1 ($\beta_\textrm{J1}^\textrm{J0805} = 4.5 \pm 0.9$) and J2 ($\beta_\textrm{J2}^\textrm{J0805} = 6.3 \pm 1.0$), are superluminal and consistent with each other within the uncertainties. The trajectory of J1 seems to follow that of J2, its position angle changes from $\sim190\degr$ to $\sim160\degr$ while travelling up to $\sim2$~mas projected distance from the core (Fig.~\ref{fig:J0805+6144-pm}). From that point on, the bending jet becomes remarkably straight up to $\sim8$~mas separation. The diffuse outer component J3 was absent in the first few years and could be reliably detected only at four later epochs. The time coverage and sampling for J3 is not sufficient for a reliable proper motion determination, therefore we give an upper limit to its apparent speed in Table~\ref{tab:pm}.

\begin{figure*}
    \centering
    \includegraphics[width=8cm]{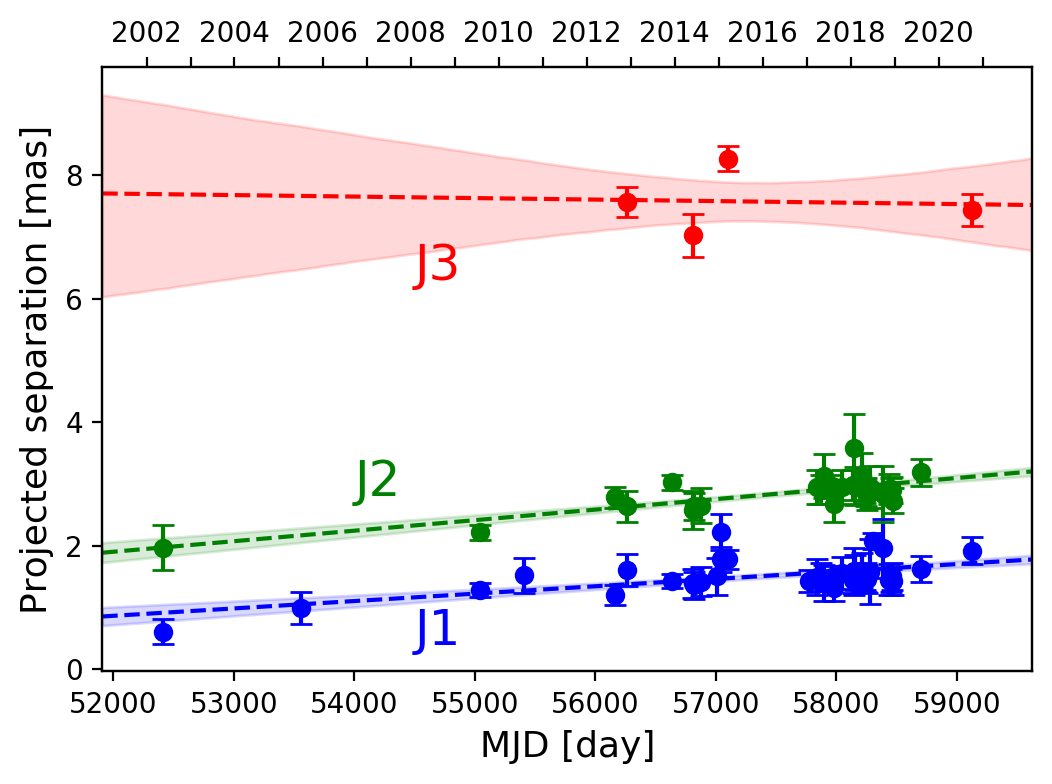}
    \includegraphics[width=8cm]{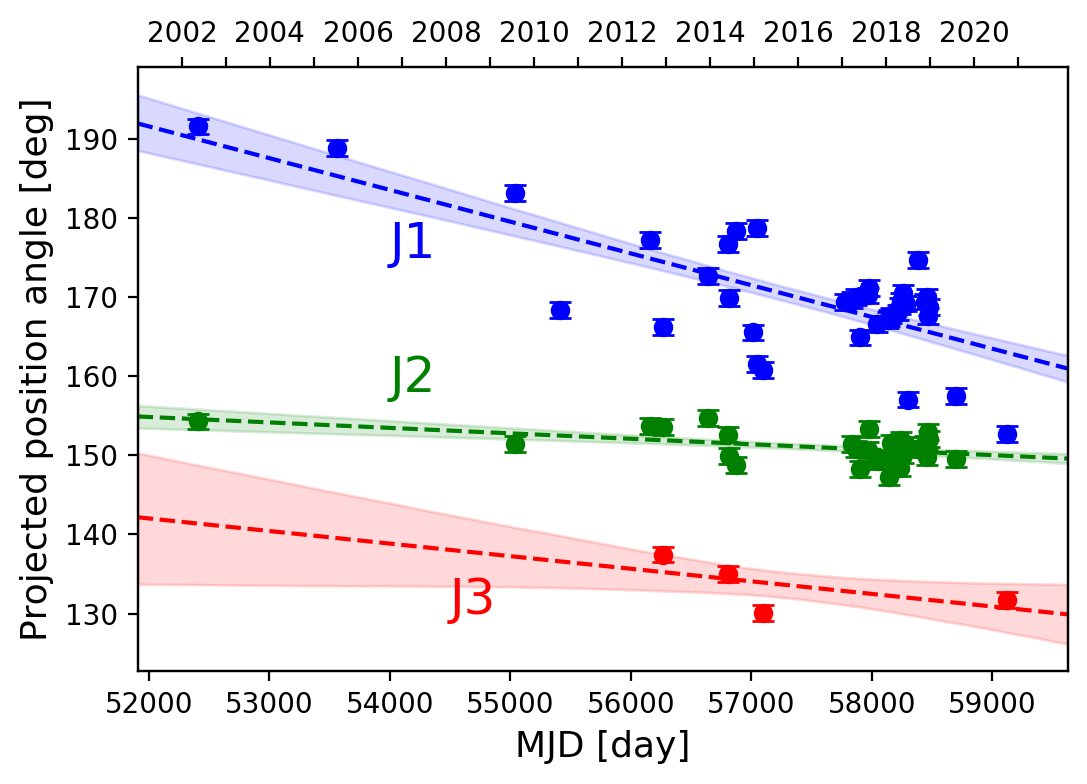}
    \includegraphics[width=8cm]{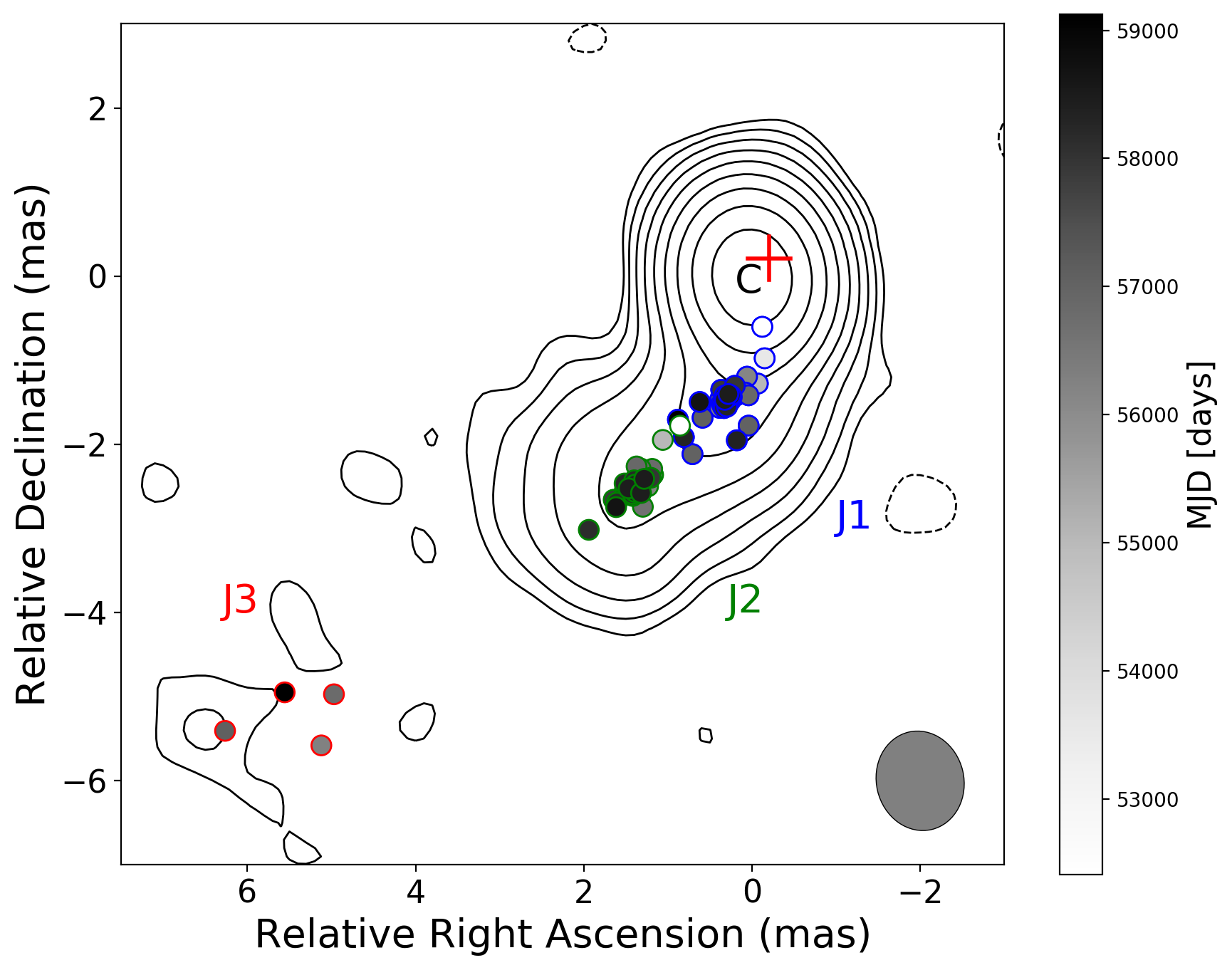}
    \caption{Proper motion plots and 8.6-GHz VLBI image of J0805$+$6144. The dashed lines represent the best-fit linear model. The shaded areas represent the $1 \sigma$ uncertainties of each fit. The colouring of the components is the following: J1 -- blue, J2 -- green, J3 -- red. \textit{Top left:} Radial proper motion of each component. \textit{Top right:} Jet component position angles as a function of time. \textit{Bottom:} Locations of the jet components plotted onto the 8.6-GHz VLBI total intensity image made on 2018 May 19 (experiment ug002h). The peak intensity is $402$~mJy\,beam$^{-1}$, with the lowest contours drawn at $\pm1.1$~mJy\,beam$^{-1}$. The positive  contour levels increase by a factor of 2. The size of the restoring beam is $1.2\,\mathrm{mas} \times 1.0\,\mathrm{mas}$ (FWHM) at $\mathrm{PA}=13\fdg3$ (measured from north through east), as indicated in the bottom-right corner.
    The shading indicates the observing time in MJD (Modified Julian Date) at the given position of each jet component. 
    The red cross marks the \textit{Gaia} DR3 optical position, its size indicates the formal uncertainties.
    } 
    \label{fig:J0805+6144-pm}
\end{figure*}

In the case of J1658$-$0739 (Fig.~\ref{fig:J1658-0739-pm}), the four jet components follow a bent trajectory, starting in the northeastern direction, and then turning northwards. The probably newly-emerging innermost components J1 and J2 were not detected at the first epoch in 2002 due to the limited angular resolution. The proper motions are apparently superluminal, except for J2 ($\beta_\textrm{J2}^\textrm{J1658} = 0.9 \pm 1.6$) which is consistent with being stationary (Table~\ref{tab:pm}). The apparent advance speed of the outer J3 and J4 components is about $14 c$. It should be noted that these outer components are in the more extended parts of the jet. It is harder to detect and accurately determine their position, resulting in higher uncertainties.

\begin{figure*}
    \centering
    \includegraphics[width=8cm]{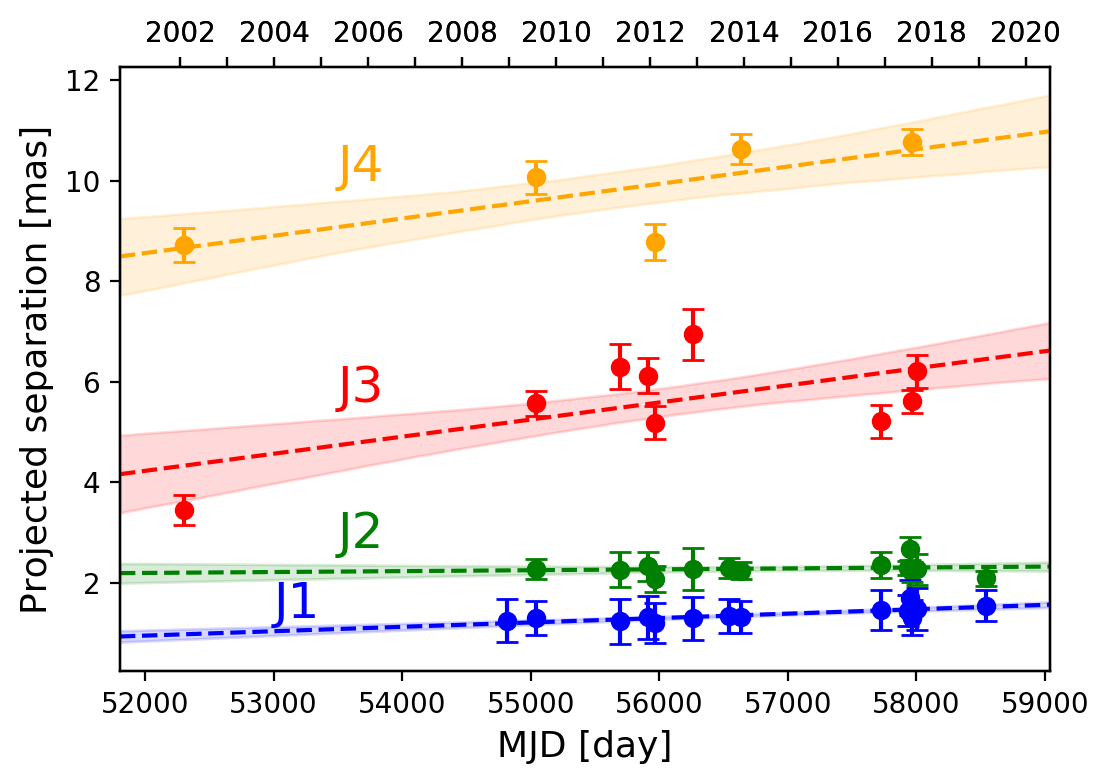}
    \includegraphics[width=8cm]{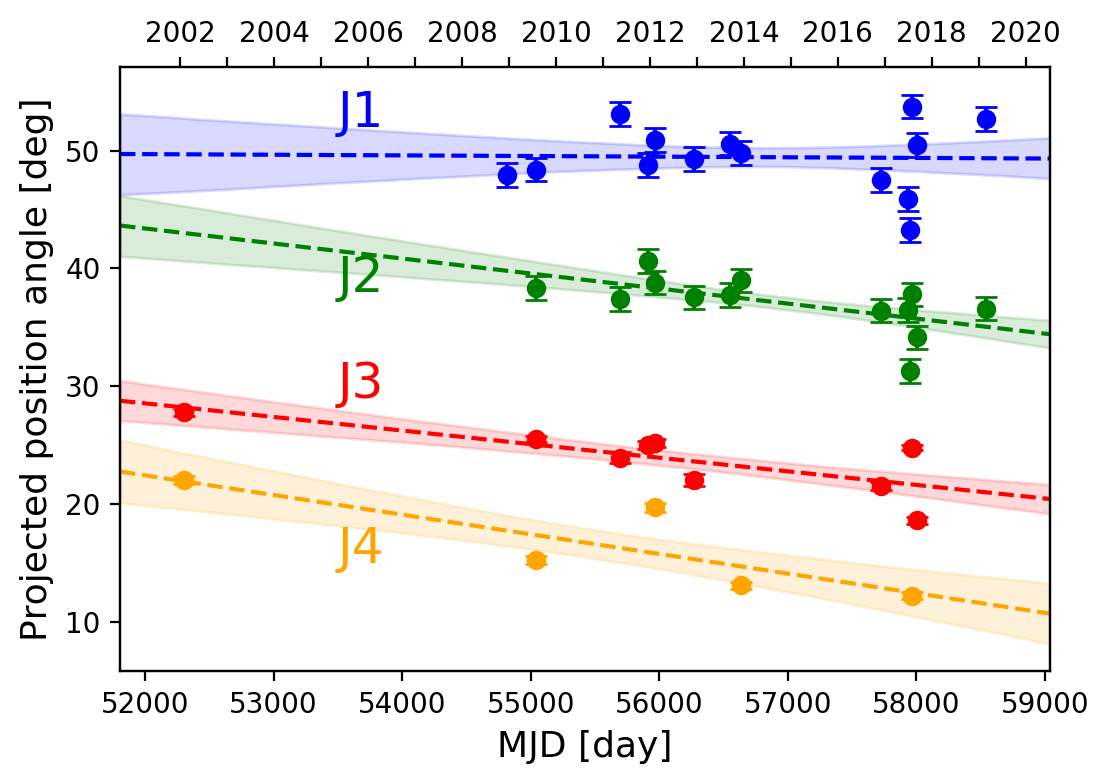}
    \includegraphics[width=8cm]{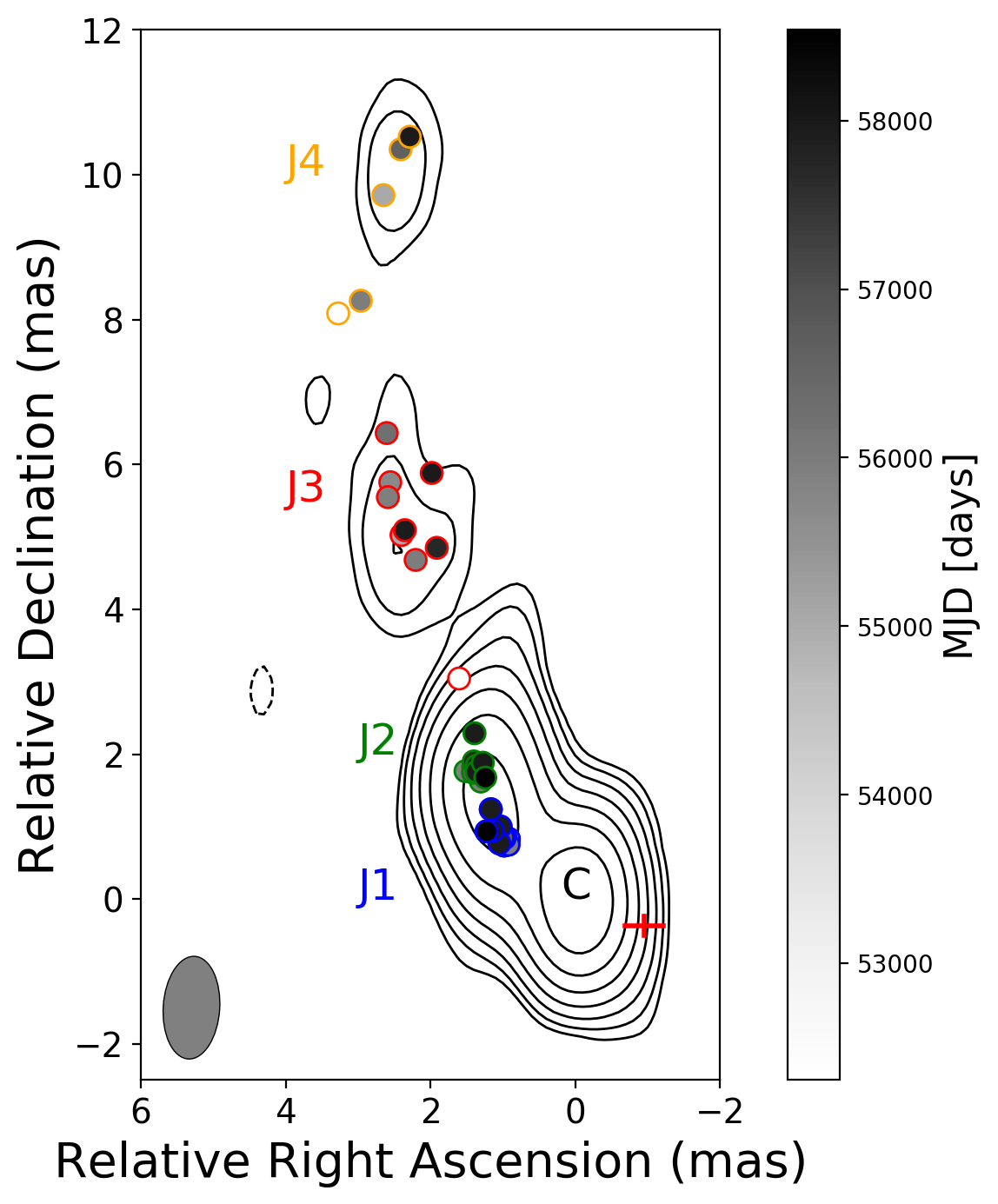}
    \caption{Proper motion plots and 8.6-GHz VLBI image of J1658$-$0739. The dashed lines represent the best-fit linear model. The shaded areas represent the $1 \sigma$ uncertainties of each fit. The colouring of the components is the following: J1 -- blue, J2 -- green, J3 -- red, J4 -- orange. \textit{Top left:} Radial proper motion of each component. \textit{Top right:} Jet component position angles as a function of time. 
    \textit{Bottom right:} Locations of the jet components plotted onto the VLBI image made on 2017 Aug 1 (experiment rv125). The peak intensity is $226$~mJy\,beam$^{-1}$, with the the lowest contours drawn at $\pm1.7$~mJy\,beam$^{-1}$. The positive contour levels increase by a factor of 2. The size of the restoring beam is $1.4\,\mathrm{mas} \times 0.8\,\mathrm{mas}$ (FWHM) at $\mathrm{PA} = -3\fdg7$, as indicated in the bottom-left corner.
    The shading indicates the observing time in MJD at the given position of each jet component. The red cross marks the \textit{Gaia} DR3 optical position, its size indicates the formal uncertainties.
} 
    \label{fig:J1658-0739-pm}
\end{figure*}

\subsection{Flux densities and brightness temperatures}
\label{subsec:phys_param}

Figures~\ref{fig:J0805+6144-plots} and \ref{fig:J1658-0739-plots} show the flux density (\textit{left}) and brightness temperature (\textit{middle}) as a function of time for J0805$+$6144 and J1658$-$0739, respectively. 

The VLBI component flux densities were determined from the Gaussian model fitting described in Section~\ref{sec:data}. We present the core flux densities, as well as the sum of the flux densities of the core and the closest jet component (J1), to better describe the innermost radio-emitting region. When the J1 jet component is still close to the core, the flux density of the compact central region may not be well represented by the core component alone. The emergence of a new jet component is often associated with a flux density outburst. For a while, C and the outward-moving J1 are blended together, as the new component cannot be distinguished from the core because of the limited angular resolution of the network. 

In the case of J0805$+$6144 (Fig.~\ref{fig:J0805+6144-plots}), the flux density in the central region is dominated by the core, and the J1 flux density stays rather constant. From this, and the generally decreasing trend in the light curve, it is likely that the outburst possibly associated with the ejection of J1 had happened prior to the start of the VLBI observations. The VLBI flux density curve of J1658$-$0739 (Fig.~\ref{fig:J1658-0739-plots}) also indicates an overall decreasing trend. However, during the first half of the monitoring period, the central (C+J1) flux density is dominated by the recently ejected jet component. As J1 advances outward, its flux density decreases, while the core flux density remains stable. The brightening of the core started again around 2012. 

The redshift-corrected brightness temperatures were calculated following the equation of \citet{1982ApJ...252..102C}:
\begin{equation} \label{eq:tb}
    T_{\rm{b}} = 1.22 \times 10^{12} \, (1 + z) \frac{S_\nu}{\phi_\mathrm{a}\phi_\mathrm{b} \nu^2} \,\, [\mathrm{K}],
\end{equation}
where $z$ is the redshift, $S_\nu$ the integrated flux density of the core in Jy, $\nu$ the observing frequency in GHz, $\phi_\mathrm{a}$ and $\phi_\mathrm{b}$ the major and minor axes (FWHM) of the fitted elliptical Gaussian in mas, respectively. \cite{2005AJ....130.2473K} give a formula for the size of the minimum resolvable source component with the interferometer. To determine $b_{\rm \psi}$ (the half-power beam width measured along an arbitrary position angle $\psi$), we followed Appendix B of \citet{2021ApJ...910..105H}. The calculated minimum resolvable size was substituted in Eq.~\ref{eq:tb} instead of $\phi_\mathrm{a}$ and $\phi_\mathrm{b}$ as an upper limit if the source was unresolved at a given epoch. This way, the brightness temperatures obtained are lower limits. 

When $T_{\rm b}$ exceeds the equipartition value, $T_\mathrm{b,eq} \approx 5 \times 10^{10}$\,K \citep{1994ApJ...426...51R}, the emission is considered relativisically enhanced (Doppler-boosted). The Doppler factors can be calculated by assuming an intrinsic brightness temperature $(T_\mathrm{b,int})$ in the source as  
\begin{equation} \label{eq:delta}
 \delta = \frac{T_\mathrm{b}}{T_\mathrm{b,int}},
\end{equation}
where $T_\mathrm{b}$ is derived from VLBI measurements at a given frequency as described above. We note that the measured brightness temperatures are variable with time (Fig.~\ref{fig:J0805+6144-plots}--\ref{fig:J1658-0739-plots}), and the intrinsic brightness temperatures may usually be lower than the equipartition value \citep[see e.g.][]{2006ApJ...642L.115H,2021ApJ...923...67H}. For this reason, we adopted the intrinsic brightness temperature, $T_\mathrm{b,int} = 4.10 \times 10^{10}$ K, from \citet{2021ApJ...923...67H}, who found $T_\mathrm{b,int}$ to be close to the equipartition value when taking the median core $T_{\rm b}$ over many observational epochs. Before calculating the Doppler factor, we estimated the median brightness temperatures ($T_\mathrm{b,median}$) based on our measured $T_\mathrm{b}$ values of the VLBI core at $7.6-8.6$~GHz (see Table \ref{tb:results}), including those that are considered as lower limits. This way, the resulting $T_\mathrm{b,median}$ values will be reasonable estimates and they are not affected much by some single $T_{\rm b}$ lower limits among the measurements.}

The radio emission of the core is Doppler boosted in both sources in most of the epochs and the median values exceed $T_{\rm b, int}$, too. The $T_\mathrm{b,median}$ for J0805$+$6144 is $ 4.59 \times 10^{11}$~K, while for J1658$-$0739, $T_\mathrm{b,median} = 1.11 \times 10^{11}$~K. We used these $T_\mathrm{b,median}$ values in the numerator of Eq.~\ref{eq:delta} to obtain a single $\delta$ value for the cores. The Doppler factors determined this way (Table~\ref{tab:pm}) can be considered characteristic to these jets.

\subsection{Lorentz factors and jet viewing angles}
\label{subsec:lorentz}

\begin{figure*}
    \centering
    \includegraphics[width=5.7cm]{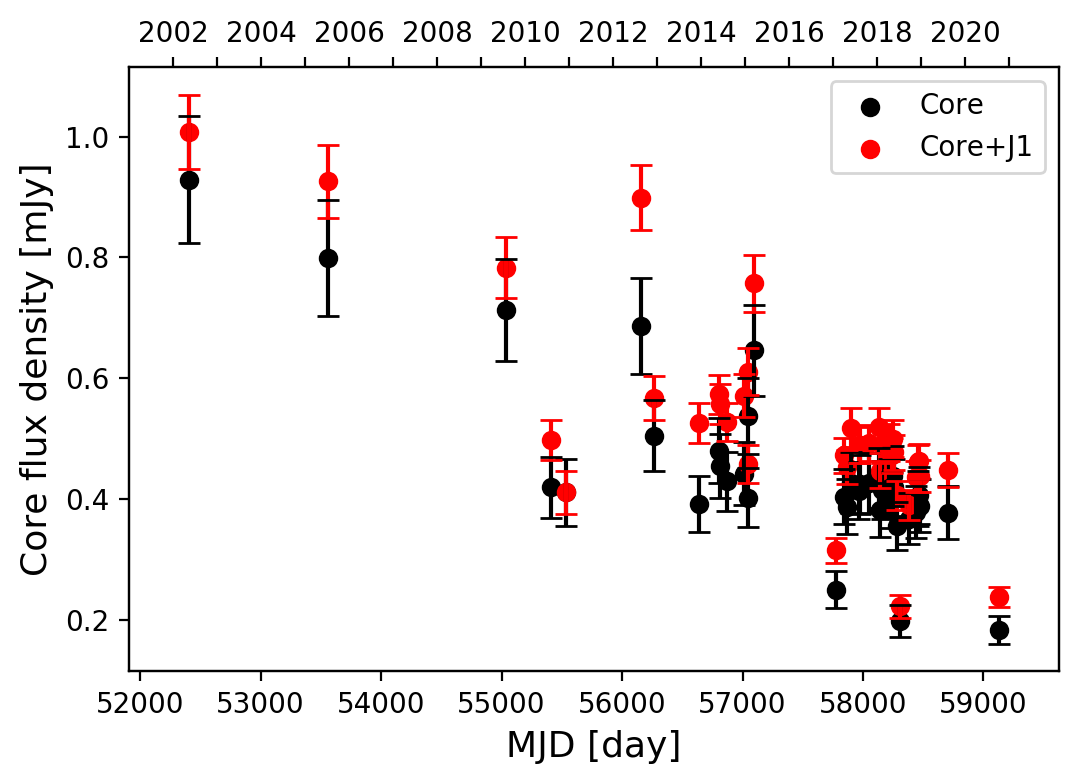}
    \includegraphics[width=5.7cm]{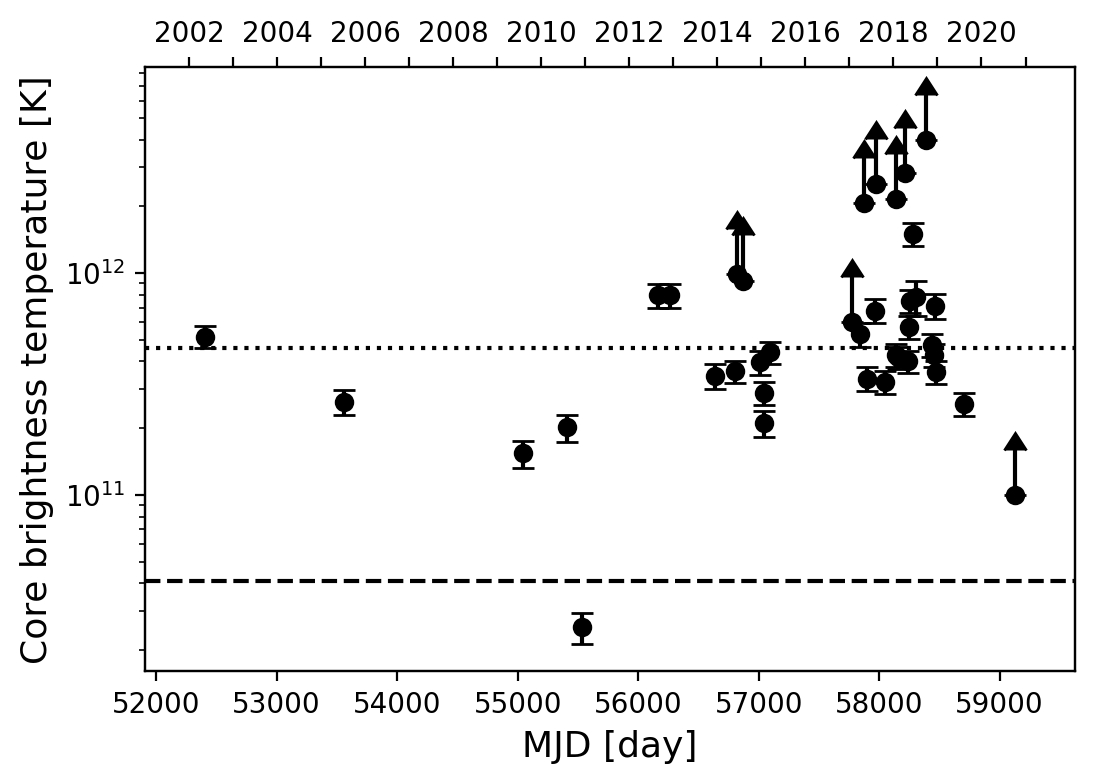}
    \includegraphics[width=6.1cm]{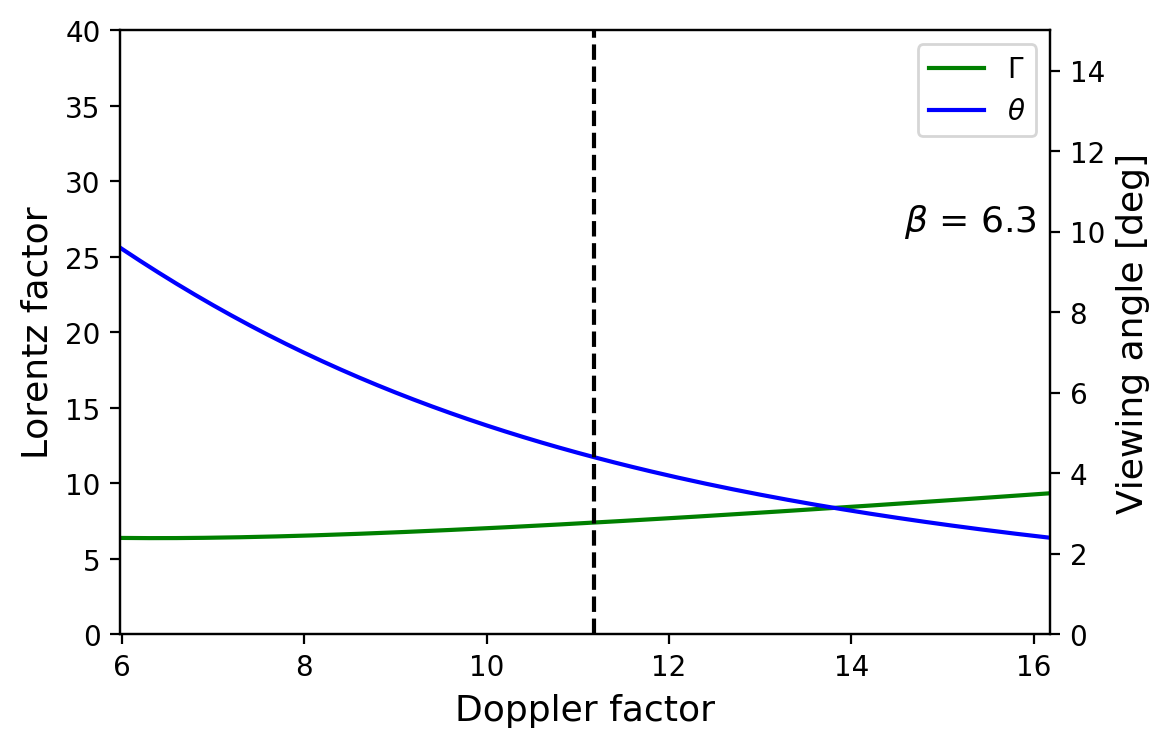}
    \caption{The core flux density, redshift-corrected brightness temperature, and jet parameter plots for J0805+6144. \textit{Left}: Changes in the core flux density during the time covered by the VLBI observations. The fitted flux densities of the core are plotted in black, the sum of the core and the J1 jet component flux densities in red. \textit{Middle}: The brightness temperature of the core as a function of time. Only the lower limit of the $T_\mathrm{b}$ is plotted if the source was unresolved with the interferometer. The dotted line shows the median brightness temperature while the dashed line indicates the intrinsic brightness temperature, $4.1 \times 10^{10}$~K, adopted from \citet{2021ApJ...923...67H}.
    \textit{Right}: The Lorentz factor (\textit{green}) and the jet viewing angle (\textit{blue}) as a function of the Doppler factor, based on the jet component with the fastest apparent speed. The dashed line corresponds to the  estimated Doppler-factor ($\delta \simeq 11.2$)}. 
    \label{fig:J0805+6144-plots}
\end{figure*}

\begin{figure*}
    \centering
    \includegraphics[width=5.8cm]{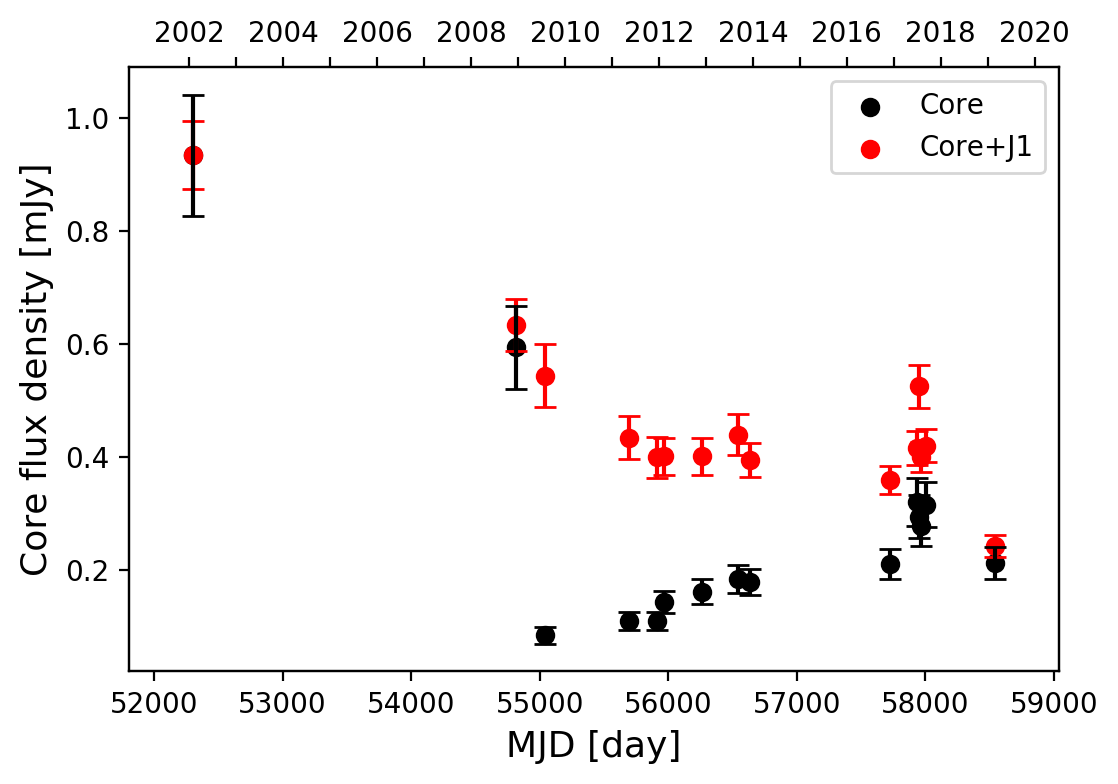}
    \includegraphics[width=5.8cm]{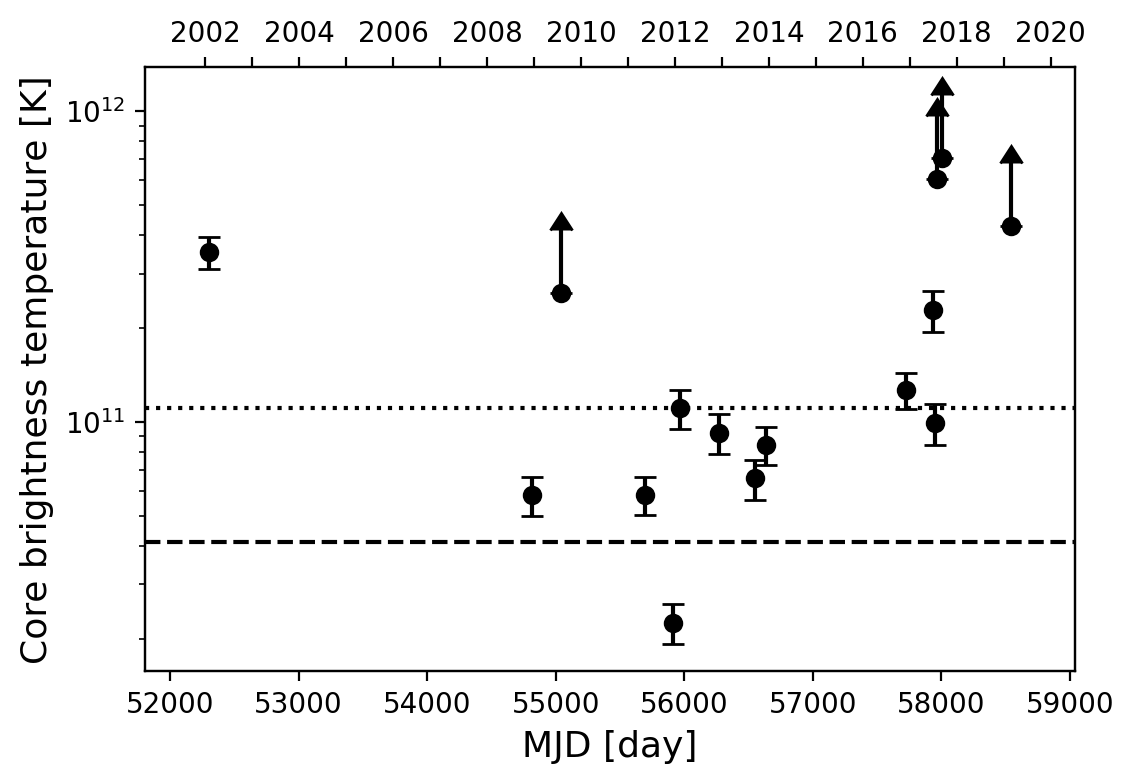}
    \includegraphics[width=6cm]{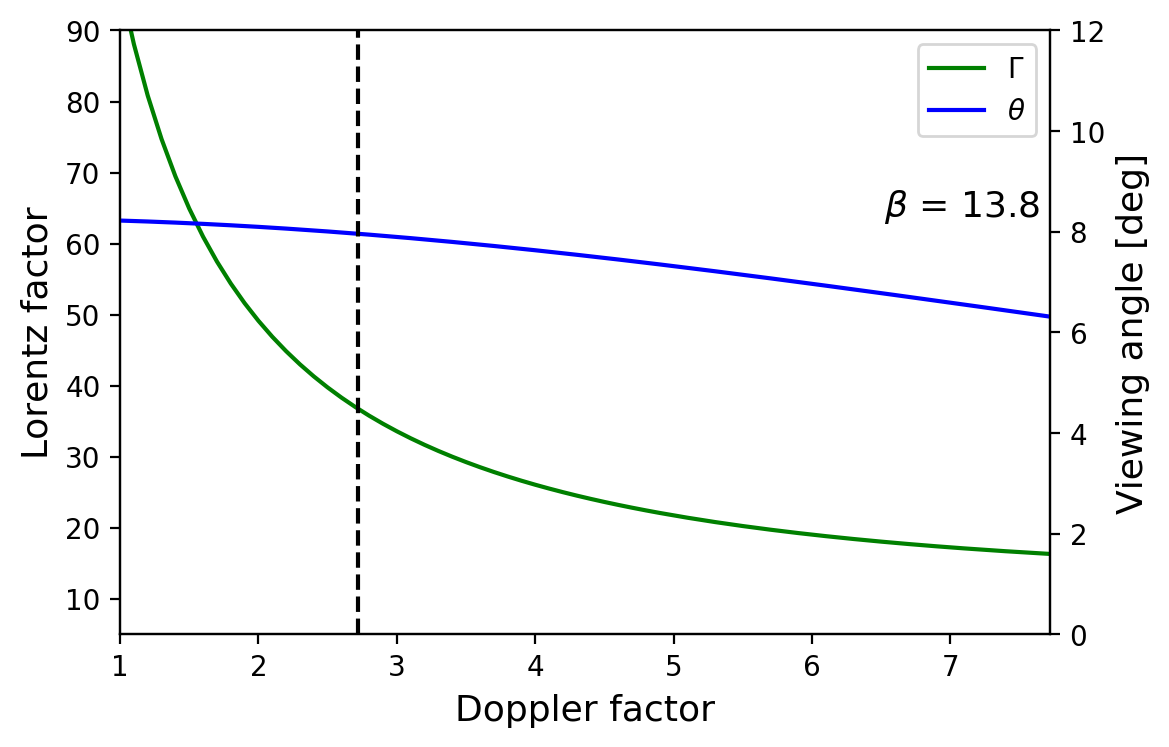}
    \caption{The core flux density, redshift-corrected brightness temperature, and jet parameter plots for J1658$-$0739. The description is the same as for Fig.~\ref{fig:J0805+6144-plots}. 
    \textit{Right}: The dashed line corresponds to the estimated Doppler-factor ($\delta \simeq 2.7$).}
    \label{fig:J1658-0739-plots}
\end{figure*}

There are two commonly used methods for determining the bulk Lorentz factor in a blazar jet. One is to fit the broad-band SED of a blazar (e.g. \citealt{1997A&A...324..395B}). The other method, also applied in this paper, is based on VLBI observations of the radio jet, using the Doppler factor and the apparent superluminal speed. The following equations from \citet{1993ApJ...407...65G} can be used to calculate the Lorentz factor and the jet viewing angle with respect to the line of sight: 
\begin{equation}
    \Gamma = \frac{\beta^2 + \delta^2 + 1}{2\delta},
\end{equation}
\begin{equation}
    \tan \theta = \frac{2\beta}{\beta^2 + \delta^2 - 1},
\end{equation}
where $\beta$ is the apparent superluminal speed of the jet component in the units of the speed of light $c$. Table~\ref{tab:pm} contains the Lorentz factors and the jet viewing angles calculated using the highest apparent jet component proper motion determined for the sources (see e.g. \citealt{2009A&A...494..527H,2022ApJ...937...19Z}). Figures~\ref{fig:J0805+6144-plots} and \ref{fig:J1658-0739-plots} (\textit{right}) show the Lorentz factors and the jet viewing angles as a function of the Doppler factor in the viccinity of the characteristic $\delta$ values. It should be noted here that, because of the assumptions made in Section~\ref{subsec:phys_param}, we cannot determine an exact value for the Lorentz factors and the viewing angles, but rather close estimates.

In the case of J0805$+$6144, the Lorentz factor is $\Gamma \approx 7.4$ and the inclination angle is $\theta \approx 4.4\degr$ (Fig.~\ref{fig:J0805+6144-plots}). The $T_{\rm b}$ lower limit at the last epoch is below the calculated $T_\mathrm{b,median}$ and its exact value could still slightly influence the resulted Doppler factor. The Lorentz factor for J1658$-$0739 is $\Gamma \approx 36.6$ (Fig.~\ref{fig:J1658-0739-plots}), a rather high value, but not unprecedented at high redshifts \citep{2022ApJ...937...19Z}. The viewing angle is $\theta \approx 8.0\degr$. Both our targets have their jet viewing angle within $\theta \approx 10\degr$ as expected for typical blazars.


\section{Discussion}
\label{sec:discussion}

\subsection{Proper motion--redshift relation for jetted quasars}
\label{subsec:prop_motion}

Distant jetted radio sources above redshift 3 are generally harder to find because of their relative weakness and the lack of bright, prominent mas-scale jet features compared to low-redshift sources. The small number of known sources \citep[e.g.][]{2017FrASS...4....9P} makes it challenging to reveal an overall picture. The few dedicated high-redshift jet proper motion studies carried out with VLBI so far, either for single or multiple objects, are e.g. \citet{2015MNRAS.446.2921F,2017MNRAS.468...69Z,2018MNRAS.477.1065P,2020MNRAS.497.2260A,2020SciBu..65..525Z,2022ApJ...937...19Z}. The cosmological time dilation requires long VLBI monitoring of the sources in the observer's frame to obtain well-sampled data to detect changes in the jet component positions. At lower redshifts, \citet{2019ApJ...874...43L} found that the distribution of bulk Lorentz factors in jets peaks between $5-15$, and it is not uncommon to reach $\Gamma \approx 40$. The high-redshift jet studies cited above found jets with $\Gamma \lesssim 40$ so far.

\citet{1988ApJ...329....1C}, \citet{1994ApJ...430..467V}, and \citet{1999NewAR..43..757K} investigated the dependence of the apparent proper motion on redshift ($\mu-z$ diagram) using growing with time samples of sources with measured kinematics of jet components. These investigations, among other factors, accounted for selection effects and underlying distribution of Lorentz factors in the jets. As indicated, $\mu-z$ dependences can be considered as inputs into cosmological tests. Later studies by \citet{2008A&A...484..119B}, \citet{2012ApJ...758...84P}, and \citet{2019ApJ...874...43L} found that
at low redshifts the relation satisfies the prediction of the concordance $\Lambda$CDM cosmological model if the majority of the jets, as measured, have $\Gamma \le 25$. According to \citet{2022ApJ...937...19Z} and references therein, this valid at high redshifts as well, and the bulk Lorentz factor $\Gamma \approx 40$ is only reached by one source so far. Our Lorentz factors for two $z>3$ blazar jets derived from the measured proper motions do not contradict to the findigs for other high-redshift sources known from the literature. For an up-to-date $\mu-z$ plot, see \citet{2022ApJ...937...19Z}.

\subsection{The Lorentz factors of J0805$+$6144 and J1658$-$0739}
\label{subsec:lor}

There are independent Lorentz factor estimates from the literature for both of our targets. For J0805$+$6144, \citet{2020ApJ...897..177P} derived $\Gamma = 14$ from SED fitting. Our $\Gamma \approx 7.4$ is lower by a factor of 2. On the other hand, from a single X-ray observation, \citet{2020MNRAS.498.2594S} obtained $\Gamma = 1.1 \pm 0.1$, which is significantly below both values. For J1658$-$0739, only one other $\Gamma$ estimate is found in the literature, again based on SED fitting by \citet{2017ApJ...851...33P}. These authors obtained $\Gamma = 10$, which is well below our $\Gamma \approx 36.6$. \citet{2017ApJ...851...33P} also gave an estimate for the jet inclination angle, $\theta = 3\degr$, in contrast to our $\theta \approx 8.0\degr$. In our model with $\beta=13.8$, their jet viewing angle would correspond to $\delta\simeq 20$.

The differences between the results from the VLBI analysis and SED fitting might be because different regions of the jet are probed by radio interferometry and X-ray measurements. Also, X-ray emission can be produced at various places, like the corona, the inner jet or large, kpc-scale lobes in AGNs. Regarding VLBI imaging, it is possible that the size of the core is overestimated, leading to the underestimation of the brightness temperatures, thus the Doppler factor, too. \citet{2017MNRAS.464.4306N} showed that it can happen when the VLBI core component is blended with a very nearby jet component and these are unresolved by the interferometer. Moreover, variability can also be a source of uncertainty when estimating physical parameters. 

\subsection{The inner region of J1658--0739}
\label{subsec:inner_j1658}

The inner jet region of J1658$-$0739 showed different faces during the period covered by the VLBI observations (Sect.~\ref{subsec:proper_motion}--\ref{subsec:phys_param}, Figs.~\ref{fig:J1658-0739-pm}, \ref{fig:J1658-0739-plots}). Notably, the J2 component has no significant proper motion in the radial direction (Table~\ref{tab:pm}). It could be a standing shock (\citealt{1948sfsw.book.....C, 1988ApJ...334..539D}) where a stationary knot can appear well-separated from the core. Standing shocks are belived to be formed where the jet recollimates or returns to higher density and pressure. Normally, the external pressure should drop far away from the central engine, causing the jet to become wider and less dense than at the time of the ejection. While this type of morphological feature is well represented at low redshifts based on jet monitoring VLBI programs \citep[e.g.][]{2017ApJ...846...98J,2021ApJ...923...30L}, they are somewhat rare in high-redshift relativistic jets which show superluminal motion. The absence of these features might be caused by the scarce sample of high-$z$ jet kinematic studies. More stationary hot spots similar to ours are found in e.g. 3C\,395 \citep{1985AJ.....90.1989W}, 4C\,39.25 \citep{1988ApJ...334..539D}, also in J0753$+$4231 \citep{2022ApJ...937...19Z} at high redshift, $z = 3.595$. An alternative explanation we should consider for the stationary J2 component is that it could be a projection effect rather than a physical standing shock. The gradual curvature observed in the jet supports the idea that the apparent stationarity of J2 is caused by a jet bending that slows down the flow speed when projected onto the sky plane. 

There are two published VLBI images of this source in the literature, taken at different frequencies. \citet{2000ApJS..131...95F} presented a 5-GHz VLBA image (angular resolution $\sim3$~mas, dynamic range $\sim1000:1$) as part of the \textit{VSOP} Prelaunch Survey. It shows a bright core component with an extension to the northeast, consistent with the structure seen in our 8.6-GHz map. 
Another, high-frequency image made at 24~GHz \citep[angular resolution $\sim1$~mas, dynamic range $\sim150:1$,][]{2010AJ....139.1713C} shows a bright component at the phase centre and a much fainter emission peak towards the southwest at a separation of $\sim$ 1 mas and a position angle $\mathrm{PA} = 225\degr$. Concerning the absolute astrometric positions of J1658$-$0739, the X-band (8.6~GHz) and K-band (24~GHz) ICRF3 solutions \citep{2020A&A...644A.159C} agree with each other within the uncertainties, while the $Gaia$ optical position is separated from them by $\sim 1$~mas in the position angle of about $220\degr$ (i.e. to the southeast, Fig.~\ref{fig:J1658-0739-pm}). Notably, the \textit{Gaia} position seems to coincide with the weak secondary component in the 24-GHz image \citep{2010AJ....139.1713C}, suggesting that this is the actual core location. The apparent discrepancy between the source structure seen in our 8.6-GHz images (a bright compact core and a jet pointing towards the northwest, Fig.~\ref{fig:J1658-0739-plots}) and the 24-GHz image of where the brightest feature is in the northwest with a weaker secondary component towards the southeast may be reconciled in the context of the flux density variability and the structural changes observed in the source. Unfortunately, we lack 8.6-GHz observations in the period $2003-2007$ when the 24-GHz image was made. However, the decreasing trend in the core+J1 flux density (Fig.~\ref{fig:J1658-0739-plots}) and the emergence of a new inner jet component in the 8.6-GHz maps starting from 2008 suggest an outburst in or shortly before 2002 when a new component was born. This could have resulted in the bright feature that appears in the 24-GHz image taken on 2007 Mar 30 \citep{2010AJ....139.1713C}. Later, as J1 moved further away from the core, the two features became clearly resolved at 8.6~GHz as well. 

\section{Summary}
\label{sec:summary}

In this paper, we presented a kinematic analysis of the radio jets in two bright high-redshift ($3<z<4$) blazars, J0805+6144 and J1658$-$0739, for the first time. We analysed 7.6--8.6-GHz VLBI data covering nearly two decades of archival observations, supplemented by EVN imaging observations in 2018 and 2019. We used imaging data with mas-scale angular resolution to model the brightness distribution of the sources. We identified multiple jet components across the observing epochs in both J0805+6144 and J1658$-$0739, and estimated their apparent proper motion. The mas-scale jet structure of J1658$-$0739 contains an apparently stationary component which might be caused by a projection effect or associated with a standing recollimation shock. By interpreting the fitted component flux densities as a function of time, a higher-resolution archival 24-GHz VLBI image \citep{2020A&A...644A.159C}, and the radio and optical absolute astrometric positions, we propose the explanation that the jet component marked by J1 might have been caused by a prominent outburst in this AGN that happened around 2000.

The measurements allowed us to obtain estimates of the characteristic physical and geometric parameters of these blazar jets, the bulk Lorentz factor and the inclination angle with respect to the line of sight. The results support the blazar nature of these sources. The derived apparent superluminal motions are ranging between $1 \lesssim \beta \lesssim 14$. Our proper motion measurements add to the sparsely sampled high-redshift part of the apparent proper motion--redshift relation. The derived Lorentz factors are consistent with values found in other $z > 3$ radio-loud AGNs, supporting expectations from the standard $\Lambda$CDM cosmological model without requiring unphysically high jet speeds. Regarding the future, larger samples of high-$z$ jets are needed to draw statistically robust conclusions. In addition, observations with longer time baselines would help verify stationary features, and multi-frequency VLBI studies could better identify core positions.

\section*{Acknowledgements}

We are grateful for the insightful comments and suggestions of the anonymous referee.
The EVN is a joint facility of independent European, African, Asian and North American radio astronomy institutes. Scientific results from data presented in this publication are derived from the following EVN project code: ET036. 
The National Radio Astronomy Observatory is a facility of the National Science Foundation operated under cooperative agreement by Associated Universities, Inc. 
The authors acknowledge the use of Astrogeo Center database maintained by L. Petrov. 
This work presents results from the European Space Agency (ESA) space mission Gaia. Gaia data are being processed by the Gaia Data Processing and Analysis Consortium (DPAC). Funding for the DPAC is provided by national institutions, in particular the institutions participating in the Gaia MultiLateral Agreement (MLA). The Gaia mission website is \url{https://www.cosmos.esa.int/gaia}. The Gaia archive website is \url{https://archives.esac.esa.int/gaia}.
This research has made use of the NASA/IPAC Extragalactic Database (NED) which is operated by the Jet Propulsion Laboratory, California Institute of Technology, under contract with the National Aeronautics and Space Administration.
We thank the Hungarian National Research, Development and Innovation Office (OTKA K134213) for support. TA is supported by the National SKA Programme of China (grant no. 2022SKA0120102), the Tianchi Talents Programme of Xinjiang Uygur Autonomous Region and the open funding of the Laboratory of Pinghu. YZ is supported by the National SKA Programme of China (grant no. 2022SKA0120102), Shanghai Sailing Program (grant no. 22YF1456100).

\section*{Data Availability}

The calibrated data underlying this article are available in the Astrogeo Center database maintained by L. Petrov at 
\url{http://astrogeo.org/cgi-bin/imdb_get_source.csh?source=J0805%2B6144} and
\url{http://astrogeo.org/cgi-bin/imdb_get_source.csh?source_name=1656-075}. The raw correlated EVN data are publicly accessible in the EVN Data Archive at \url{http://archive.jive.nl/scripts/portal.php}. The  calibrated EVN visibility data and images underlying this article will be shared on reasonable request to the corresponding author.



\bibliographystyle{mnras}
\bibliography{main} 

\bsp (
\label{lastpage}
\end{document}